\documentclass[reprint,amsmath,amssymb, aps,prl]{revtex4-1}

\usepackage{graphicx}% Include figure files
\usepackage{dcolumn}% Align table columns on decimal point
\usepackage{bm}% bold math
\usepackage{epstopdf}
\usepackage{lineno}

\begin{document}
\title{{Machine Learning Inverse Problem for Topological Photonics}}

\author{Laura Pilozzi}%
\email{laura.pilozzi@isc.cnr.it}
\affiliation{Institute for Complex Systems, National Research Council (ISC-CNR), Via dei Taurini 19, 00185 Rome, Italy}%
\author{Francis A. Farrelly}
\affiliation{Institute for Complex Systems, National Research Council (ISC-CNR), Via dei Taurini 19, 00185 Rome, Italy}%
\author{Giulia Marcucci}%
\affiliation{Institute for Complex Systems, National Research Council (ISC-CNR), Via dei Taurini 19, 00185 Rome, Italy}%
\affiliation{Department of Physics, University Sapienza, Piazzale Aldo Moro 5, 00185 Rome, Italy}%
\author{Claudio Conti}
\affiliation{Institute for Complex Systems, National Research Council (ISC-CNR), Via dei Taurini 19, 00185 Rome, Italy}%
\affiliation{Department of Physics, University Sapienza, Piazzale Aldo Moro 5, 00185 Rome, Italy}%
\email{claudio.conti@uniroma1.it}

\date{\today}

\graphicspath{{Images/}}

\begin{abstract}
Topological concepts open many new horizons for photonic devices,
from integrated optics to lasers~\cite{Lu,wu,carusotto2018}.
The complexity of large scale topological devices asks for an effective solution of the inverse problem: how best to engineer the topology for a specific application? We introduce a novel machine learning approach to the topological inverse problem~\cite{bishop,duda,murphy}.
We train a neural network system with the band structure of the Aubry-Andre-Harper model and then adopt the network for solving the inverse problem.
Our application is able to identify the parameters of a complex topological insulator in order to obtain protected edge states at target frequencies.
One challenging aspect is handling the multivalued branches of the direct problem and discarding unphysical solutions. We overcome this problem by adopting a self-consistent method to only select physically relevant solutions.
We demonstrate our technique in a realistic topological laser design and by resorting to the widely available open-source TensorFlow library~\cite{TF}.
Our results are general and scalable to thousands of topological components. This new inverse design technique based on machine learning potentially extends the applications of topological photonics, for example, to frequency combs, quantum sources, neuromorphic computing and metrology.
\end{abstract}


\maketitle
The rapidly growing interest in topological photonics~\cite{Lu,wu} is leading to the design of complex structures for the many applications of optical topological insulators.\cite{carusotto2018} One leading goal of topological photonics is photon transport protected from unwanted random scattering. 
This is achieved by realizing analogues of the quantum Hall effect~\cite{Haldane,Raghu,Wang} through magnetic-like Hamiltonians in photonic systems~\cite{Rechtsman}. In the optical domain, topological insulators~\cite{Hasan} have been implemented in modulated honeycomb lattices~\cite{Rechtsman}, in arrays of coupled optical-ring resonators~\cite{Hafezi} and optical quantum walks~\cite{kitagawa}. Geometry-independent topological structures have been proposed to obtain nonreciprocal single mode lasing~\cite{Bahari,Bandreseaar4005,Hararieaar4003,amo2017} as well as systems with balanced gain and loss for parity-time symmetric structures with topological order~\cite{Rivolta,pilo}. Emulations of four-dimensional physics have also been reported \cite{zilberberg_photonic_2018,lohse_exploring_2018}.

One challenge in this field is to find an effective methodology for the inverse problem in which the target optical properties result from topological characteristics. 
Although various computational techniques are available, these require specific implementations tailored to the task at hand. 
Machine learning (ML)~\cite{bishop,duda,murphy} has recently been proposed as an encompassing technology for dealing with greatly differing problems through a unified approach. 
ML techniques have shown a remarkable growth in sophistication and application scope in multiple fields~\cite{zde,carra,zhang}; ML offers exciting perspectives in topological photonics.
ML is applied in two main classes of problems: (i) classification for categorizing information and (ii) regression to predict continuous values from supervised training.  Unlike parametric regression - in which a best fit of the data is determined on the basis of a specific function - ML regression employs a neural network (NN) emulating the behaviour of the data on which it has been trained: {\it ``the NN learns the model"}.

In this paper, we employ ML regression for solving the inverse problem in topological photonics.
We apply advanced ML techniques to design photonic topological insulators enabling innovative applications through custom tailoring of desired optical parameters. 
In our approach, we introduce a twist in order to ensure that only physically possible solutions are found. 
This twist is based on a self-consistent cycle in which a tentative solution obtained from the inverse problem NN is run through the direct problem NN in order to ensure that the solution obtained is indeed viable. 

\begin{figure*}[t]
       	\includegraphics[width=2.0\columnwidth]{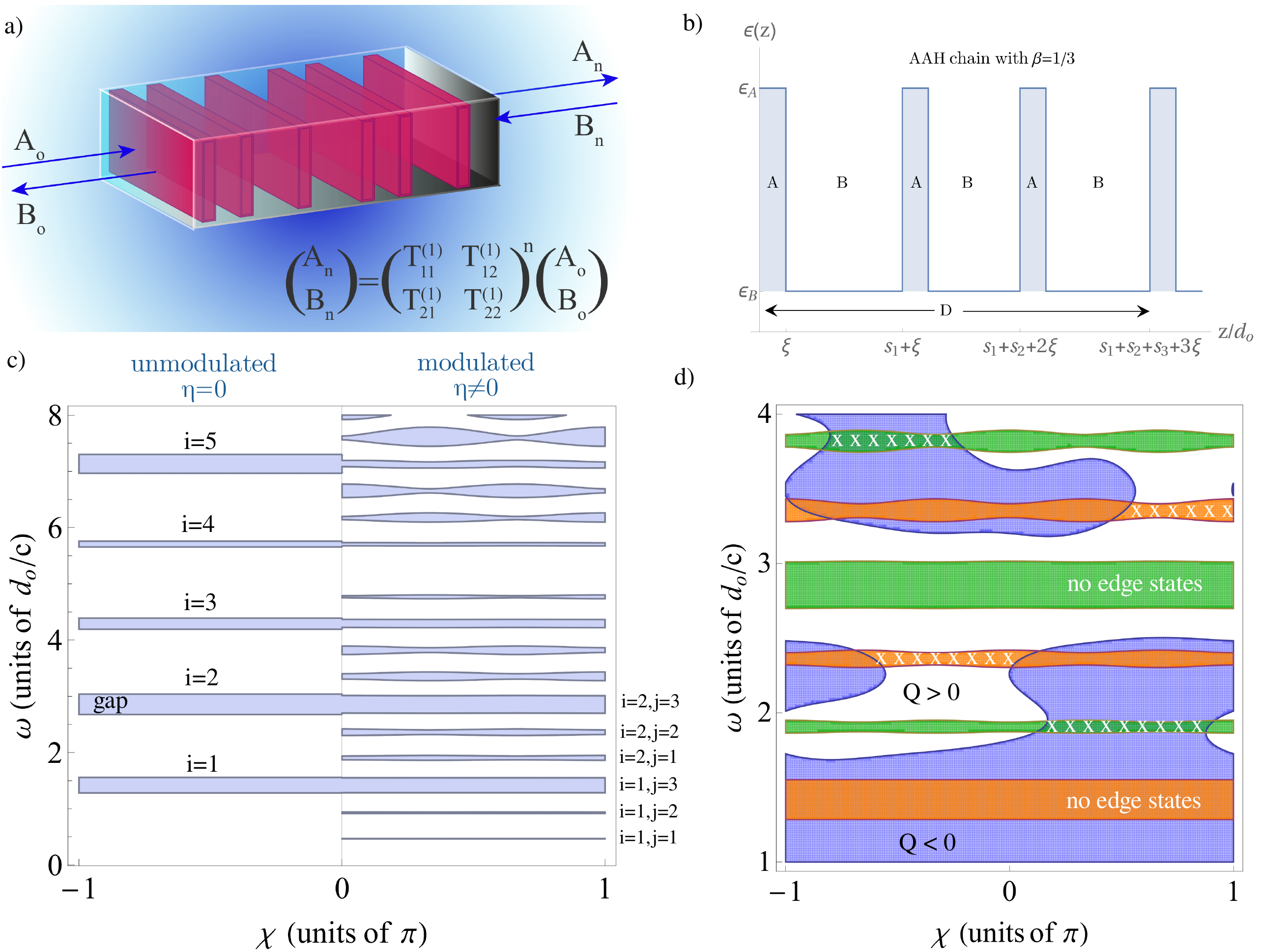}
	\caption{(Color online) a) Scheme of the topological optical structure. b) Dielectric function profile for an AAH chain with  $\beta=1/3$, with $s_i=[z_{i+1}-z_i-L_A]/d_o$. c) Band diagram with $\chi=\phi+\pi(2\beta-1)/2$.  For $|\chi/\pi|>1$ one can identify the gaps of the unmodulated structure (blue regions). The range $|\chi /\pi| < 1$ shows the gaps 
with Harper modulation: each gap of the unmodulated structures ($|\chi/\pi|>1$) splits into $q$ bands. d) Orange and green regions correspond to gaps. White areas indicate the regions where $Q(\omega,\chi,\xi)>0$, blue the regions where $Q(\omega,\chi,\xi)<0$. Edge states are possible only in the regions with crosses in orange and green gaps. \label{fig:f1}}
\end{figure*}

We consider one of the simplest structures that support non-trivial topological properties.
In one-dimensional (1D) systems, synthetic magnetic fields occur by lattice modulation~\cite{Kraus} of the optical structure. In the Aubry-Andre-Harper (AAH) model~\cite{Aubry,Harper}, identical sites - resonators, two-level atoms, waveguides, etc. - are centered at positions $z_n = d_o\left( {n + \eta \delta_n^H } \right)$,  with $n$ an integer label, $d_o$ the primary lattice period, $\eta$ the modulation strength and $\delta_n^H = \cos (2\pi\beta n + \phi )$ the Harper modulation~\cite{Harper}. The parameter $\beta$ is the frequency of the Harper modulation. Together, $\beta$ and the phase shift $\phi$ furnish the topological properties by a ``two-dimensional ancestor'' mapping ~\cite{Posha}.  
The 2D ancestor is characterized by the dependence of the dielectric function on the coordinate $z$ and on the parameter $\phi$, which acts as a periodic artificial coordinate.
Hence the phase $\phi$ can be treated as a wave vector in a fictitious auxiliary direction~\cite{Posha}.
For $\beta=p/q$ with $p>0$ and $q>0$ integers, the lattice displays two commensurate periods with $q$ sites $z_n$ in the unit-cell.
Properly chosen parameters give rise to nontrivial topological phases with protected states at the border of the structure. 
These ``edge-states'' are hallmarks of topological insulators. The phase $\phi$ tunes edge-state eigenfrequency in the photonic band-gaps.

Our photonic topological insulator is an array of layers $A$ of normalized thickness $\xi=L_A/d_o$, centered in $z_n$, in an homogeneous bulk of material $B$.
This kind of structure can be effectively modeled by the transfer matrix technique~\cite{Chew,pilo},  as reported in Fig.~\ref{fig:f1}a. In this figure $A_0$ and $A_n$ are the initial and final amplitudes of the right-travelling waves; while $B_0$ and $B_n$ are their equivalent for the left-travelling wave amplitudes. 
As detailed in Methods, we obtain the transfer matrix for the single period $T^{(1)} (\omega, \phi, \xi )$ with elements $T_{11}^{(1)}$, $T_{12}^{(1)}$, $T_{21}^{(1)}$ and $T_{22}^{(1)}$. Fig.~\ref{fig:f1}a shows the final wave amplitudes $A_n,B_n$ by the $n$-fold repeated action of $T^{(1)} (\omega, \phi, \xi )$ on $A_0,B_0$. The dielectric constant profile - for the case $\beta=1/3$ - is schematically illustrated in Fig.~\ref{fig:f1}b.

For $\eta= 0$, we have a periodical unmodulated structure with frequency bandgaps labeled by an integer $i$. For $\eta\neq 0$, each gap of the unmodulated structure splits into $q$ gaps, each one labelled by indices $(i,j)$ 
($j=1,...,q$)~\cite{Hofs}. This splitting is shown in Fig.~\ref{fig:f1}c for $\beta=1/3$ with respect to the variable $\chi=\phi+\pi(2\beta-1)/2$.

As detailed in Methods and illustrated in Fig.~\ref{fig:f1}d, enforcing 
boundary conditions at the left edge~\cite{Hatsugai,Tauber} and defining the function $Q(\omega, \phi, \xi)$ enables one to establish the presence of edge states corresponding to poles $\omega_t$ of the reflection coefficient.
However, the function $\omega_t=\omega(\chi, \xi)$ 
cannot be analytically inverted to express the geometrical parameters $\chi$ and $\xi$ in terms of the variable $\omega_t$.
Exploiting ML techniques we solve this inverse problem and design topological insulators with target edge modes.

\begin{figure*}[t]
	\includegraphics[width=\textwidth]{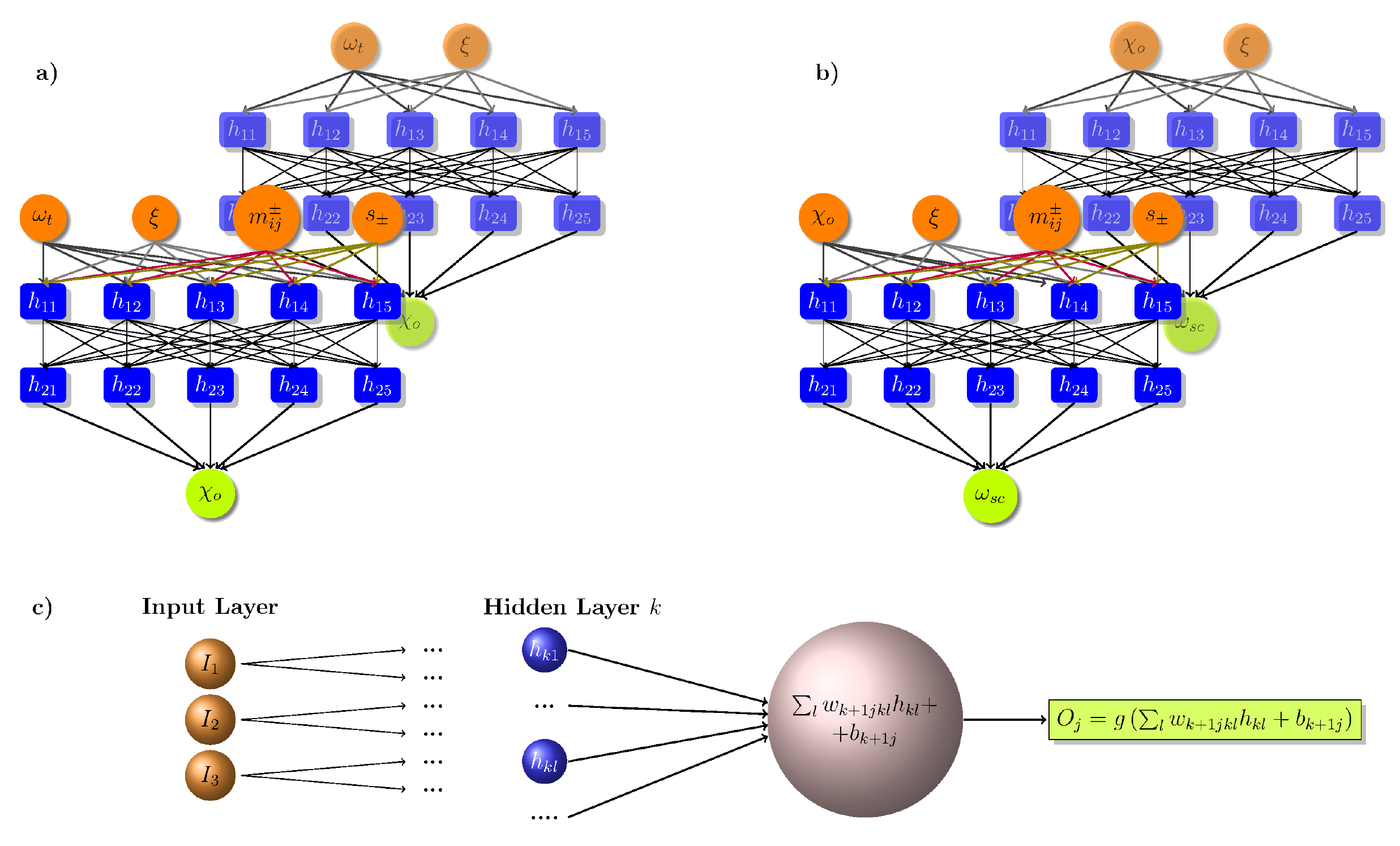}
	\caption{(Color online) Architecture of fully-connected feed-forward neural networks. Orange and green circles are the input and output units, respectively. Blue ones represent the nodes of the hidden layers. Interconnections among the units are given by arrows. The networks in the background are specific to the unfolded problem; in the foreground we show the networks with extra mode and trend inputs. a) Inverse problem network. b) Direct problem network. c) Single unit scheme. The node performs a linear combination of its inputs followed by a nonlinear activation function. \label{fig:fnet}}
\end{figure*}

The inverse problem in artificial NN theory - and therefore in ML - is widely discussed in numerical modelling, engineering and other fields~\cite{kabir, Gosal, Aoad}. Regression in ML optimizes a NN so that a given vector input ($\mathbb{R}^n$) will result in a scalar ($\mathbb{R}$) output, 
emulating the behaviour of the training data. A regressive NN is a configuration of computational layers such that a specific set of input nodes $\underline{I}$ is connected to a single output node, through a configurable set of $N_h$ hidden layers each containing $n_i$ nodes $h_{ij}$, where $i=1, ... N_h$ and $j=1, ... n_i$.
Examples of such regressive NNs are shown in Figs.~\ref{fig:fnet}a,b.
A generic node $k+1,j$, shown in part c, receiving as inputs $h_{ki}$, with $i=1,...,n_k$, yields on output $h_{k+1j}=g\left(\sum_l w_{k+1 j k l}h_{k l}+b_{k+1 j}\right)$, with $g(x)$ being a nonlinear activation function, $w_{k+1 j k l}$ the weight of $h_{kl}$ on $h_{k+1 j}$ with a bias term $b_{k+1j}$. Following accepted practice, our activation function is $g(x)=\mathrm{tanh}(x)$.

Optimization of the NN is performed by minimizing a cost function by a gradient descent method that updates weights and biases. In the initial state,  weights $w_{ijkl}$ are selected from a truncated normal and biases are set to zero.
Training applies this procedure to a data-set randomly split into two separate classes: (i) an actual training set and (ii) a validation set. The network is iteratively updated until the error on the validating data-set converges to a given rate. 

The inverse topological problem at hand is to obtain the desired optical behaviour: a target edge-state at frequency $\omega_t$, which is an input to the design (Fig.~\ref{fig:fnet}a). 
ML techniques achieve this result by modeling the multidimensional nonlinear relationships among all the structure parameters $\omega_t$, $\chi$, $\beta$, $\epsilon_A$, $\epsilon_B$ and $\xi$. In our specific case, the data-set fixes $\epsilon_A$, $\epsilon_B$, $\beta$ at the values $\epsilon_A=9$, $\epsilon_B=4$ and $\beta=1/3$.

\begin{figure*}[t]
	\includegraphics[width=\textwidth]{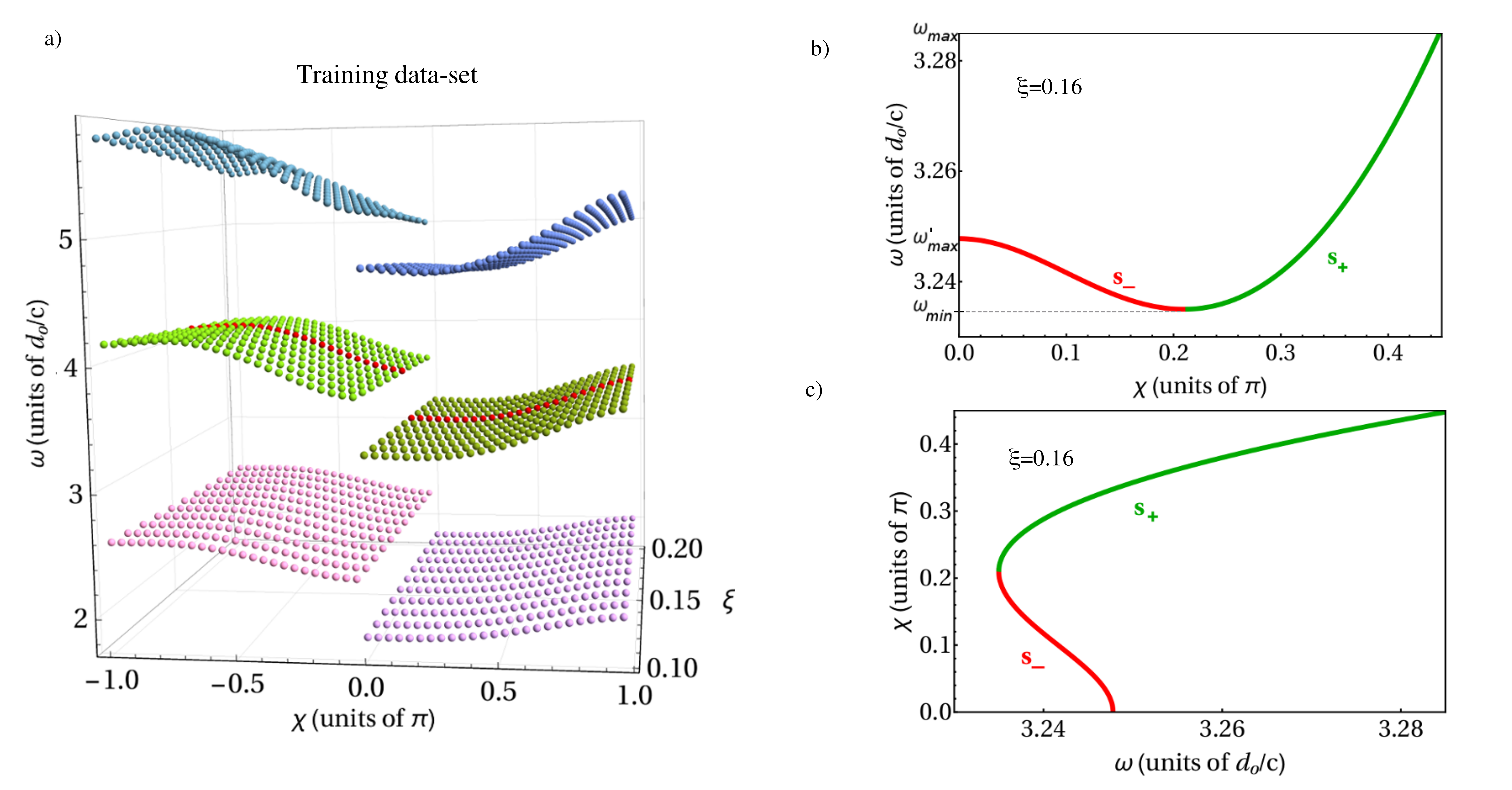}
	\caption{(Color online) a) The training dataset. Points are the real component (mode frequencies) of the complex roots of the function $T^{(1)}_{12}(\omega, \chi, \xi)$. b) Edge state dispersion for a specific mode and $\xi$ value, exhibiting a positive $s_+$ (green) and negative $s_-$ (red) trend. c) Multivalued relationship of features and targets for the same  edge mode dispersion. The $s_\pm$ labels are used for training the inverse model.\label{fig:f2}}
\end{figure*}

First we generate a data-set to train our NNs by numerically computing the complex roots of $T_{12}^{(1)} (\omega, \chi, \xi)$ covering the region interest for parameters $\chi$ and $\xi$. The real part of these roots, shown in Fig~\ref{fig:f2}a, represents the edge states dispersion.
Interestingly the same data-set can be used both for the inverse and direct NN training phase, by suitably selecting the features and target fields.
The inverse problem NN (Fig.~\ref{fig:fnet}a) targets a value $\chi=\chi_o$, 
a topological parameter on the basis of features including $\omega_t$. 
For a direct problem (Fig.~\ref{fig:fnet}b) the mode frequency $\omega_t$ would be the target of a network whose features include the topological parameters ($\chi, \xi$).

The data-set contains various branches since there exist an edge state for each band gap (i,j) with $j\neq3$, as results by Eq.~(\ref{eq:nu}) in Methods.
Due to the folding of the Brillouin zones, the edge state frequency $\omega(\chi,\xi)$ is then a multi-mode function, which we unfold by introducing a label $m_{ij}^{\pm}$ for each mode; here $i=1, ... \infty$ and $j=1, ... q$, while the sign $\pm$ indicates modes in the positive/negative $\chi$ domain.  
In Fig.~\ref{fig:f2}a, data points with different $ij$ values are identified with different colors and, solving the inverse problem is a matter of determining when these surfaces intercept a specific target value of the $\omega$ axis. 
Three outcomes are possible: a single value for $\chi$ and $\xi$ when a monotonic mode surface is intercepted, no solution for values of $\omega$ laying between surfaces, and multiple solutions in other cases.
This implies that the feature set ($\chi$, $\xi$, $m^\pm_{ij}$) is insufficient.
To tackle this problem we take into account the trend $s_{\pm}=\mathrm{sgn}\left(d\omega_t / d\chi\right)$ as an additional variable. The NNs with this enlarged feature set are illustrated in  Figs.~\ref{fig:fnet}a,b.

In the terminology used in ML the mode index $m^\pm_{ij}$ and trend $s_\pm$ labels are {\it categorical features} and lead to two possible courses of action for the actual implementation of the NNs used in our problem. One in which a single NN is constructed in a hybrid feature space with both continuous variables (real valued $\xi$'s and $\chi$'s) and categorical features, as illustrated in Fig.~\ref{fig:fnet}b. Another course is to adopt multiple independent NNs, one NN for each mode and each trend.

The single NN approach is hindered by the presence of discontinuities in the features domain, as evidenced in Fig.~\ref{fig:f2}a, so we have chosen to use multiple independent NNs.

Moreover, when considering the solution provided by the inverse NNs, we identify a specific problem in the use of ML as they may furnish solutions that are not physical.
An example of this issue is given in Fig. \ref{fig:f2}b where - for a fixed band and a fixed $\xi$ - the curve representing $\omega$ as a function of $\chi$ is shown together with its inverse (inset \ref{fig:f2}c).
Inverting the function $\omega(\chi)$, we consider an interval of values for $\omega$ spanning from its minimum $\omega_{min}$ to the maximum $\omega_{max}$, but for the two branches of the inverse function $\chi(\omega)$ - identified by colors in inset c - the range of $\omega$ is different. For example, for the red branch the maximal value of $\omega$ is $\omega_{max}'<\omega_{max}$.
When the target frequency is outside of this range, the NN produces an output outside of the physically acceptable range for $\chi$.
The inverse NN can furnish spurious non-physical solutions.

Our approach tackles this issue by a two-step self-consistent cycle: (i) in the first stage a desired input $\omega_t$ forms part of the feature set ($\omega_t, m^\pm_{ij}, s_\pm$) resulting in the output $\chi_o$ of the inverse NN; this set is used as input ($\chi_o, m^\pm_{ij}, s_\pm$) to a direct problem network; (ii) in the second stage, the target of this direct network $\omega_{sc}$ is compared with the input value $\omega_t$ and  $\chi_o$ is retained as a solution of the inverse model if $|\omega_{sc}-\omega_t| < \delta$ with $\delta$ is a user-defined small positive quantity. The value of $\delta$ affects the model accuracy. A reasonable choice can be $\delta \sim E_j^{max}$ (with j=I,D), i.e., the maximum value of the squared error functions for the inverse (I) and the direct (D) networks.
\begin{figure*}[t]
\begin{center}
	\includegraphics[width=\textwidth]{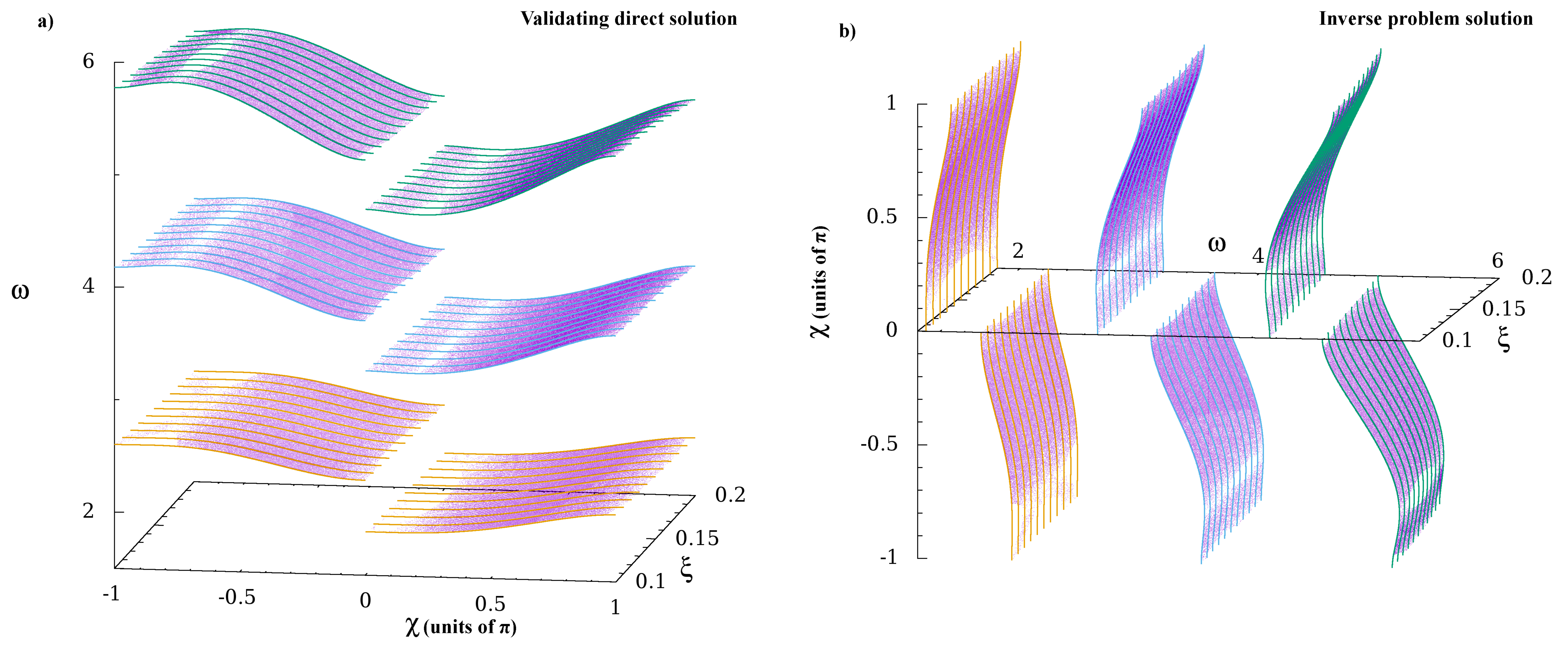}
	\caption{(Color online) Reconstruction of edge states dispersion by NN models. a) Direct problem solution as reproduce by our self-consistent cycle. b) Inverse problem solution. $\omega$ is in units of $d_0/c$. \label{fig:ftrid}}
\end{center}
\end{figure*}
\\
The training dataset was generated with eleven sets of $\xi$ ranging from $0.10$ to $0.20$ in steps of $0.01$ and for each set $\chi$ spans $-\pi$ to $\pi$ with 997 equally spaced values.
Results based on using an array of NNs each composed of 5 hidden layer of 131 nodes are shown in Fig.~\ref{fig:ftrid} together with its training set (colored lines). The model was developed using 80\% of the dataset randomly chosen the rest being used for validation and comprising of 250,000 steps. The purple dots in this figure are based on 100 values of $\xi$ while exploring the $\omega$ domain with a resolution of $10^{-5}$. Each array element is trained for a specific value of the categorical features and pertains to either the positive or the negative $\chi$ domain.  

The results of applying the direct and inverse NNs, portrayed in Fig.~\ref{fig:ftrid}a and b respectively, show that the proposed method gives accurate solutions matching the original data in the whole range of interest.
Figure ~\ref{fig:ftrid} clearly shows that our machine learning strategy solves the inverse topological design problem.
\section*{Discussion}
The inverse problem in topological design is solved by a supervised machine learning regression technique. We employ a self-consistent procedure to rule out unphysical solutions enabling tailored engineering of protected edge-states.
We successfully tackle multivalued functions introducing categorical features, as the trend, which tags training data according to their gradient's sign. Discontinuous domains are effectively treated by adopting multiple independent neural networks each one specific to its domain.
Our general method can be extensively applied - well beyond the example considered in this work - and may also be exploited for other physical systems in topological science, as polaritonics \cite{Kartashov, Mihalache}, quantum technologies and ultra-cold atoms \cite{PhysRevX.7.031057,Mancini1510}. 
The method is scalable to very complex structures involving hundreds of topological devices, as those recently considered for large scale synchronization \cite{PartoSegev}, and frequency comb generation \cite{Laura}, eventually including non-hermitian systems \cite{Longhi2018,Zeuner}. Further applications include 2D and 3D topological systems \cite{Bahari} and quantum sources and simulations \cite{zilberberg_photonic_2018,lohse_exploring_2018}.
\section*{Methods}
\textit{TensorFlow ---}
Tensorflow is Google's versatile open-source multiplatform dataflow library capable of efficiently performing machine learning tasks such as implementing neural networks. Multidimensional data arrays, referred to as ``tensors'' are executed on the basis of stateful dataflow graphs, hence the name TensorFlow. For our final code implementation Tensorflow version 1.3 with python API bindings was used.
 
The nature of our problem is such that there is a discontinuity in $\xi = 0$ which cannot be correctly handled by a single NN bridging this point; this is relevant to both the inverse and direct cases. Breaking up the data-set into two parts to be used for two separate NNs is the simplest solution to this problem. 

Another interesting aspect is related to the fact that the feature set in our inverse and direct NNs contain both continuous and discrete variables. The discrete variables can either be treated as such or handled by constructing multiple NNs each relative to a specific value of the discrete variable. The trend variable which has two possible values is one such case as is the mode number. In our code we have implemented a flexible system which allows one to decide which discrete variables are to be included in each NN, the others being broken up into arrays of NNs one for each value of the variable. Once the bookkeeping issues have been tackled this generalized approach allows one to tailor the problem to the given data-set. 

\textit{Transfer matrix ---} Given the stepped and periodic dielectric function of period $D=qd_o$:
\[
\varepsilon _\phi  (z) = \left\{ \begin{array}{l}
 \varepsilon _A \quad z_n  - L_A /2 \le z \le z_n  + L_A /2 \\ 
 \varepsilon _B \quad z_n  + L_A /2 \le z \le z_{n + 1}  - L_A /2 \\ 
 \end{array} \right.
\]
in each layer the electric field can be represented as the superposition of a left- and a right-traveling wave.
Applying the boundary conditions, the matrices 
\[
M_{\alpha \gamma }  = \frac{{q_\gamma   + q_\alpha  }}{{2q_\gamma  }}\left( {\begin{array}{*{20}c}
   1 & {r_{\alpha \gamma } }  \\
   {r_{\alpha \gamma } } & 1  \\
\end{array}} \right)
\]
with $\alpha, \gamma$ = A or B and $
r_{\alpha \gamma }  = \frac{{q_\gamma   - q_\alpha  }}{{q_\gamma   + q_\alpha  }}$, describe the light propagation through the interfaces,
having introduced $q_{\alpha}=(\omega/c)\sqrt{\epsilon_\alpha}$, while the propagation within each layer A and B is given by:
\[
T_A  = \left( {\begin{array}{*{20}c}
   {e^{iq_A d_o\xi } } & 0  \\
   0 & {e^{ - iq_A d_o\xi } }  \\
\end{array}} \right), T_{B_n} = \left( {\begin{array}{*{20}c}
   {e^{iq_B d_os_n } } & 0  \\
   0 & {e^{ - iq_B d_os_n } }  \\
\end{array}} \right)
\]
where $s_n  =[z_{n + 1}  - z_n  - L_A]/d_o$ are the normalized thicknesses of the B layers.

From these we obtain the transfer matrix for the single period $
T^{(1)}(\omega)$, the matrix connecting the fields in the left side of the elementary cell to the ones in the right side:
\[
T^{(1)}  = \prod\limits_{i = 0}^{q  - 1} {T_{B(q  - i)} M} 
\]
with $M = M_{AB} T_A M_{BA}$.
The quantity $\rho=-\frac{1}{2}TrT^{(1)} (\omega, \phi,\xi )$ allows one to locate bulk bands in the regions where $\rho^2\leqslant 1$,  and gaps where $\rho^2>1$. Alternatively, the amplitude $\left| {r_\infty  (\omega, \phi, \xi )} \right|^2 $ of the reflection coefficient of the structure~\cite{Posha}
\begin{equation}
r_\infty  (\omega, \phi, \xi  ) = \frac{{e^{ik(\omega )D}  - T_{11}^{(1)} (\omega, \phi, \xi )}}{{T_{12}^{(1)} (\omega, \phi, \xi)}}, \label{eq:r_inf}
\end{equation}
where $e^{ik(\omega )D}$ is an eigenvalue of the matrix $T_{}^{(1)} (\omega, \phi,\xi )$, can also be used to locate the gaps of the system.

\textit{Band structure of the unmodulated system ---}
The unmodulated structure ($\eta=0$) features stopbands at ${\tilde \omega} _0=\omega _0 d_0/c =\pi /(\sqrt {\varepsilon _{A} } +(1-\xi)\sqrt {\varepsilon _{B} })$, where $\xi=L_A/d_o$ is the characteristic size ratio. 

\textit{$Q(\omega,\phi,\xi)$ function---} To determine the existence of the edge states one needs to specify the boundary conditions on each edge of the structure. 
For the left edge this condition is given by:
\[0=(q_b+q_a)A_1+(q_b-q_a)B_1\] 
where $A_1$ and $B_1$ are the amplitudes of the right and left-travelling waves in the first layer of the structure. 
This condition can be reformulated as \[det(b_1,a_1)=0\] with $b_1=((q_a-q_b),(q_a+q_b))^T$ and $a_1=(A_1,B_1)^T$, and together with the eigenvalues $\lambda_\pm$ and eigenvectors $v_\pm=(T_{12}^{(1)},\lambda_\pm-T_{11}^{(1)})$ of the transfer matrix $T^{(1)}$ it is possible to determine existence and dispersion of edge states.

Following \cite{Hatsugai,Tauber} it can be in fact shown that a proportionality relation exists between the boundary vector $b_1$ and the eigenvectors $v_\pm$ of the transfer matrix. So the condition for the existence of the edge states is given by $det(b_1,v_\pm)=0$ in a gap where $|\lambda_\pm|<1$.
This entails searching for the zeros of the function $F_{l,\pm}=(q_A-q_B)(\lambda_\pm-T_{11}^{(1)})-T_{12}^{(1)}(q_A+q_B)$.

Specifically, the real part of $F_{l,\pm}=0$ yields the function $Q(\omega,\phi,\xi)=Re\{T_{12}^{(1)}(q_A+q_B)-(q_A-q_B)(T_{22}^{(1)}-T_{11}^{(1)})/2\}$ and, as shown in Fig.~\ref{fig:f1}c, this implies that edge states exist only in the gaps where $|\rho|>1$ and $Q(\omega,\phi,\xi)\cdot\rho>0$. At the same time, edge states cannot exist in gaps where $Q(\omega,\phi,\xi)$ does not change sign.
Moreover, due to a bulk-boundary correspondence \cite{Graf}, the number of these edge modes is equal to the modulus of the associated topological invariant $|\nu_{ij}|$, given by the winding number of the reflection coefficient:
\begin{equation}
\nu_{ij}  = \frac{1}{{2\pi i}}\int\limits_{ - \pi }^\pi  {d\chi \frac{{\partial ln(r_\infty(\omega ,\chi ))}}{{\partial \chi }}},
\label{eq:nu}
\end{equation}
i.e., the extra phase (divided by $2\pi$) of $r_\infty(\omega,\chi)$ when $\chi$ varies in the range ($-\pi,\pi$) with $\omega$ in the stop band\cite{PoshaAr}.
\section*{Acknowledgements}
We acknowledge support from the Templeton foundation (grant number 58277), the PRIN2015 NEMO project (2015KEZNYM grant), the H2020 QuantERA project QUOMPLEX (grant number 731473), the Italian MAE project NECST.
We thank Dr. Alexander Poshakinskiy for the fruitful comments regarding the training dataset generation.
\section*{Author contribution}
C.C. conceived the initial idea and supervised the project. F.F. expanded the concept and developed the code. L.P., G.M. and C.C. developed the theoretical part. F.F. and L.P. carried out the simulations. F.F., L.P. and G.M. contribute to data analysis and figure preparation. All the authors contributed to the manuscript writing. 

%\bibliography{biblio}

\begin{thebibliography}{46}%
	\makeatletter
	\providecommand \@ifxundefined [1]{%
		\@ifx{#1\undefined}
	}%
	\providecommand \@ifnum [1]{%
		\ifnum #1\expandafter \@firstoftwo
		\else \expandafter \@secondoftwo
		\fi
	}%
	\providecommand \@ifx [1]{%
		\ifx #1\expandafter \@firstoftwo
		\else \expandafter \@secondoftwo
		\fi
	}%
	\providecommand \natexlab [1]{#1}%
	\providecommand \enquote  [1]{``#1''}%
	\providecommand \bibnamefont  [1]{#1}%
	\providecommand \bibfnamefont [1]{#1}%
	\providecommand \citenamefont [1]{#1}%
	\providecommand \href@noop [0]{\@secondoftwo}%
	\providecommand \href [0]{\begingroup \@sanitize@url \@href}%
	\providecommand \@href[1]{\@@startlink{#1}\@@href}%
	\providecommand \@@href[1]{\endgroup#1\@@endlink}%
	\providecommand \@sanitize@url [0]{\catcode `\\12\catcode `\$12\catcode
		`\&12\catcode `\#12\catcode `\^12\catcode `\_12\catcode `\%12\relax}%
	\providecommand \@@startlink[1]{}%
	\providecommand \@@endlink[0]{}%
	\providecommand \url  [0]{\begingroup\@sanitize@url \@url }%
	\providecommand \@url [1]{\endgroup\@href {#1}{\urlprefix }}%
	\providecommand \urlprefix  [0]{URL }%
	\providecommand \Eprint [0]{\href }%
	\providecommand \doibase [0]{http://dx.doi.org/}%
	\providecommand \selectlanguage [0]{\@gobble}%
	\providecommand \bibinfo  [0]{\@secondoftwo}%
	\providecommand \bibfield  [0]{\@secondoftwo}%
	\providecommand \translation [1]{[#1]}%
	\providecommand \BibitemOpen [0]{}%
	\providecommand \bibitemStop [0]{}%
	\providecommand \bibitemNoStop [0]{.\EOS\space}%
	\providecommand \EOS [0]{\spacefactor3000\relax}%
	\providecommand \BibitemShut  [1]{\csname bibitem#1\endcsname}%
	\let\auto@bib@innerbib\@empty
	%</preamble>
	\bibitem [{\citenamefont {Lu}\ \emph {et~al.}(2014)\citenamefont {Lu},
		\citenamefont {Joannopoulos},\ and\ \citenamefont {Soljai}}]{Lu}%
	\BibitemOpen
	\bibfield  {author} {\bibinfo {author} {\bibfnamefont {L.}~\bibnamefont
			{Lu}}, \bibinfo {author} {\bibfnamefont {J.~D.}\ \bibnamefont
			{Joannopoulos}}, \ and\ \bibinfo {author} {\bibfnamefont {M.}~\bibnamefont
			{Soljai}},\ }\href@noop {} {\bibfield  {journal} {\bibinfo  {journal} {Nat.
				Phot.}\ }\textbf {\bibinfo {volume} {8}},\ \bibinfo {pages} {821} (\bibinfo
		{year} {2014})}\BibitemShut {NoStop}%
	\bibitem [{\citenamefont {Wu}\ \emph {et~al.}(2017)\citenamefont {Wu},
		\citenamefont {Li}, \citenamefont {Hu}, \citenamefont {Ao}, \citenamefont
		{Zhao},\ and\ \citenamefont {Gong}}]{wu}%
	\BibitemOpen
	\bibfield  {author} {\bibinfo {author} {\bibfnamefont {Y.}~\bibnamefont
			{Wu}}, \bibinfo {author} {\bibfnamefont {C.}~\bibnamefont {Li}}, \bibinfo
		{author} {\bibfnamefont {X.}~\bibnamefont {Hu}}, \bibinfo {author}
		{\bibfnamefont {Y.}~\bibnamefont {Ao}}, \bibinfo {author} {\bibfnamefont
			{Y.}~\bibnamefont {Zhao}}, \ and\ \bibinfo {author} {\bibfnamefont
			{Q.}~\bibnamefont {Gong}},\ }\href@noop {} {\bibfield  {journal} {\bibinfo
			{journal} {Adv. Opt. Mater.}\ }\textbf {\bibinfo {volume} {5}},\ \bibinfo
		{pages} {1700357} (\bibinfo {year} {2017})}\BibitemShut {NoStop}%
	\bibitem [{\citenamefont {{Ozawa}}\ \emph {et~al.}(2018)\citenamefont
		{{Ozawa}}, \citenamefont {{Price}}, \citenamefont {{Amo}}, \citenamefont
		{{Goldman}}, \citenamefont {{Hafezi}}, \citenamefont {{Lu}}, \citenamefont
		{{Rechtsman}}, \citenamefont {{Schuster}}, \citenamefont {{Simon}},
		\citenamefont {{Zilberberg}},\ and\ \citenamefont
		{{Carusotto}}}]{carusotto2018}%
	\BibitemOpen
	\bibfield  {author} {\bibinfo {author} {\bibfnamefont {T.}~\bibnamefont
			{{Ozawa}}}, \bibinfo {author} {\bibfnamefont {H.~M.}\ \bibnamefont
			{{Price}}}, \bibinfo {author} {\bibfnamefont {A.}~\bibnamefont {{Amo}}},
		\bibinfo {author} {\bibfnamefont {N.}~\bibnamefont {{Goldman}}}, \bibinfo
		{author} {\bibfnamefont {M.}~\bibnamefont {{Hafezi}}}, \bibinfo {author}
		{\bibfnamefont {L.}~\bibnamefont {{Lu}}}, \bibinfo {author} {\bibfnamefont
			{M.}~\bibnamefont {{Rechtsman}}}, \bibinfo {author} {\bibfnamefont
			{D.}~\bibnamefont {{Schuster}}}, \bibinfo {author} {\bibfnamefont
			{J.}~\bibnamefont {{Simon}}}, \bibinfo {author} {\bibfnamefont
			{O.}~\bibnamefont {{Zilberberg}}}, \ and\ \bibinfo {author} {\bibfnamefont
			{I.}~\bibnamefont {{Carusotto}}},\ }\href@noop {} {\bibfield  {journal}
		{\bibinfo  {journal} {ArXiv e-prints}\ } (\bibinfo {year} {2018})},\ \Eprint
	{http://arxiv.org/abs/1802.04173} {arXiv:1802.04173 [physics.optics]}
	\BibitemShut {NoStop}%
	\bibitem [{\citenamefont {Bishop}(2006)}]{bishop}%
	\BibitemOpen
	\bibfield  {author} {\bibinfo {author} {\bibfnamefont {C.}~\bibnamefont
			{Bishop}},\ }\href@noop {} {\emph {\bibinfo {title} {Pattern recognition and
				machine learning}}}\ (\bibinfo  {publisher} {Springer},\ \bibinfo {year}
	{2006})\BibitemShut {NoStop}%
	\bibitem [{\citenamefont {Duda}(2001)}]{duda}%
	\BibitemOpen
	\bibfield  {author} {\bibinfo {author} {\bibfnamefont {R.}~\bibnamefont
			{Duda}},\ }\href@noop {} {\emph {\bibinfo {title} {Pattern Classification}}}\
	(\bibinfo  {publisher} {Wiley},\ \bibinfo {year} {2001})\BibitemShut
	{NoStop}%
	\bibitem [{\citenamefont {Murphy}(2012)}]{murphy}%
	\BibitemOpen
	\bibfield  {author} {\bibinfo {author} {\bibfnamefont {K.}~\bibnamefont
			{Murphy}},\ }\href@noop {} {\emph {\bibinfo {title} {Machine learning: A
				Probabilistic Perspective}}}\ (\bibinfo  {publisher} {The MIT Press},\
	\bibinfo {year} {2012})\BibitemShut {NoStop}%
	\bibitem [{TF()}]{TF}%
	\BibitemOpen
	\href@noop {} {}\bibinfo {note} {Http://tensorflow.org}\BibitemShut {NoStop}%
	\bibitem [{\citenamefont {Haldane}\ and\ \citenamefont
		{Raghu}(2008)}]{Haldane}%
	\BibitemOpen
	\bibfield  {author} {\bibinfo {author} {\bibfnamefont {F.~D.~M.}\
			\bibnamefont {Haldane}}\ and\ \bibinfo {author} {\bibfnamefont
			{S.}~\bibnamefont {Raghu}},\ }\href@noop {} {\bibfield  {journal} {\bibinfo
			{journal} {Phys. Rev. Lett.}\ }\textbf {\bibinfo {volume} {100}},\ \bibinfo
		{pages} {013904} (\bibinfo {year} {2008})}\BibitemShut {NoStop}%
	\bibitem [{\citenamefont {Raghu}\ and\ \citenamefont {Haldane}(2008)}]{Raghu}%
	\BibitemOpen
	\bibfield  {author} {\bibinfo {author} {\bibfnamefont {S.}~\bibnamefont
			{Raghu}}\ and\ \bibinfo {author} {\bibfnamefont {F.~D.~M.}\ \bibnamefont
			{Haldane}},\ }\href@noop {} {\bibfield  {journal} {\bibinfo  {journal} {Phys.
				Rev. A}\ }\textbf {\bibinfo {volume} {78}},\ \bibinfo {pages} {033834}
		(\bibinfo {year} {2008})}\BibitemShut {NoStop}%
	\bibitem [{\citenamefont {Wang}\ \emph {et~al.}(2008)\citenamefont {Wang},
		\citenamefont {Chong}, \citenamefont {Joannopoulos},\ and\ \citenamefont
		{Soljacic}}]{Wang}%
	\BibitemOpen
	\bibfield  {author} {\bibinfo {author} {\bibfnamefont {Z.}~\bibnamefont
			{Wang}}, \bibinfo {author} {\bibfnamefont {Y.~D.}\ \bibnamefont {Chong}},
		\bibinfo {author} {\bibfnamefont {J.~D.}\ \bibnamefont {Joannopoulos}}, \
		and\ \bibinfo {author} {\bibfnamefont {M.}~\bibnamefont {Soljacic}},\
	}\href@noop {} {\bibfield  {journal} {\bibinfo  {journal} {Phys. Rev. Lett.}\
		}\textbf {\bibinfo {volume} {100}},\ \bibinfo {pages} {013905} (\bibinfo
		{year} {2008})}\BibitemShut {NoStop}%
	\bibitem [{\citenamefont {Rechtsman}\ \emph {et~al.}(2013)\citenamefont
		{Rechtsman}, \citenamefont {Zeuner}, \citenamefont {Plotnik}, \citenamefont
		{Lumer}, \citenamefont {Podolsky}, \citenamefont {Dreisow}, \citenamefont
		{Nolte}, \citenamefont {Segev},\ and\ \citenamefont {Szameit}}]{Rechtsman}%
	\BibitemOpen
	\bibfield  {author} {\bibinfo {author} {\bibfnamefont {M.~C.}\ \bibnamefont
			{Rechtsman}}, \bibinfo {author} {\bibfnamefont {J.~M.}\ \bibnamefont
			{Zeuner}}, \bibinfo {author} {\bibfnamefont {Y.}~\bibnamefont {Plotnik}},
		\bibinfo {author} {\bibfnamefont {Y.}~\bibnamefont {Lumer}}, \bibinfo
		{author} {\bibfnamefont {D.}~\bibnamefont {Podolsky}}, \bibinfo {author}
		{\bibfnamefont {F.}~\bibnamefont {Dreisow}}, \bibinfo {author} {\bibfnamefont
			{S.}~\bibnamefont {Nolte}}, \bibinfo {author} {\bibfnamefont
			{M.}~\bibnamefont {Segev}}, \ and\ \bibinfo {author} {\bibfnamefont
			{A.}~\bibnamefont {Szameit}},\ }\href@noop {} {\bibfield  {journal} {\bibinfo
			{journal} {Nature}\ }\textbf {\bibinfo {volume} {496}},\ \bibinfo {pages}
		{196} (\bibinfo {year} {2013})}\BibitemShut {NoStop}%
	\bibitem [{\citenamefont {Hasan}\ and\ \citenamefont {Kane}(2010)}]{Hasan}%
	\BibitemOpen
	\bibfield  {author} {\bibinfo {author} {\bibfnamefont {M.~Z.}\ \bibnamefont
			{Hasan}}\ and\ \bibinfo {author} {\bibfnamefont {C.~L.}\ \bibnamefont
			{Kane}},\ }\href@noop {} {\bibfield  {journal} {\bibinfo  {journal} {Rev.
				Mod. Phys.}\ }\textbf {\bibinfo {volume} {82}},\ \bibinfo {pages} {3045}
		(\bibinfo {year} {2010})}\BibitemShut {NoStop}%
	\bibitem [{\citenamefont {Hafezi}\ \emph {et~al.}(2013)\citenamefont {Hafezi},
		\citenamefont {Mittal}, \citenamefont {Fan}, \citenamefont {Migdall},\ and\
		\citenamefont {Taylor}}]{Hafezi}%
	\BibitemOpen
	\bibfield  {author} {\bibinfo {author} {\bibfnamefont {M.}~\bibnamefont
			{Hafezi}}, \bibinfo {author} {\bibfnamefont {S.}~\bibnamefont {Mittal}},
		\bibinfo {author} {\bibfnamefont {J.}~\bibnamefont {Fan}}, \bibinfo {author}
		{\bibfnamefont {A.}~\bibnamefont {Migdall}}, \ and\ \bibinfo {author}
		{\bibfnamefont {J.}~\bibnamefont {Taylor}},\ }\href@noop {} {\bibfield
		{journal} {\bibinfo  {journal} {Nat. Photon.}\ }\textbf {\bibinfo {volume}
			{7}},\ \bibinfo {pages} {1001} (\bibinfo {year} {2013})}\BibitemShut
	{NoStop}%
	\bibitem [{\citenamefont {Kitagawa}\ \emph {et~al.}(2012)\citenamefont
		{Kitagawa}, \citenamefont {Broome}, \citenamefont {Fedrizzi}, \citenamefont
		{Rudner}, \citenamefont {Berg}, \citenamefont {Kassal}, \citenamefont
		{Aspuru-Guzik}, \citenamefont {Demler},\ and\ \citenamefont
		{White}}]{kitagawa}%
	\BibitemOpen
	\bibfield  {author} {\bibinfo {author} {\bibfnamefont {T.}~\bibnamefont
			{Kitagawa}}, \bibinfo {author} {\bibfnamefont {M.~A.}\ \bibnamefont
			{Broome}}, \bibinfo {author} {\bibfnamefont {A.}~\bibnamefont {Fedrizzi}},
		\bibinfo {author} {\bibfnamefont {M.~S.}\ \bibnamefont {Rudner}}, \bibinfo
		{author} {\bibfnamefont {E.}~\bibnamefont {Berg}}, \bibinfo {author}
		{\bibfnamefont {I.}~\bibnamefont {Kassal}}, \bibinfo {author} {\bibfnamefont
			{A.}~\bibnamefont {Aspuru-Guzik}}, \bibinfo {author} {\bibfnamefont
			{E.}~\bibnamefont {Demler}}, \ and\ \bibinfo {author} {\bibfnamefont {A.~G.}\
			\bibnamefont {White}},\ }\href@noop {} {\bibfield  {journal} {\bibinfo
			{journal} {Nat. Commun.}\ }\textbf {\bibinfo {volume} {3}},\ \bibinfo {pages}
		{882} (\bibinfo {year} {2012})}\BibitemShut {NoStop}%
	\bibitem [{\citenamefont {Bahari}\ \emph {et~al.}(2017)\citenamefont {Bahari},
		\citenamefont {Ndao}, \citenamefont {Vallini}, \citenamefont {Amili},
		\citenamefont {Fainman},\ and\ \citenamefont {Kant\'e}}]{Bahari}%
	\BibitemOpen
	\bibfield  {author} {\bibinfo {author} {\bibfnamefont {B.}~\bibnamefont
			{Bahari}}, \bibinfo {author} {\bibfnamefont {A.}~\bibnamefont {Ndao}},
		\bibinfo {author} {\bibfnamefont {F.}~\bibnamefont {Vallini}}, \bibinfo
		{author} {\bibfnamefont {A.~E.}\ \bibnamefont {Amili}}, \bibinfo {author}
		{\bibfnamefont {Y.}~\bibnamefont {Fainman}}, \ and\ \bibinfo {author}
		{\bibfnamefont {B.}~\bibnamefont {Kant\'e}},\ }\href@noop {} {\bibfield
		{journal} {\bibinfo  {journal} {Science}\ }\textbf {\bibinfo {volume}
			{358}},\ \bibinfo {pages} {636} (\bibinfo {year} {2017})}\BibitemShut
	{NoStop}%
	\bibitem [{\citenamefont {Bandres}\ \emph {et~al.}(2018)\citenamefont
		{Bandres}, \citenamefont {Wittek}, \citenamefont {Harari}, \citenamefont
		{Parto}, \citenamefont {Ren}, \citenamefont {Segev}, \citenamefont
		{Christodoulides},\ and\ \citenamefont {Khajavikhan}}]{Bandreseaar4005}%
	\BibitemOpen
	\bibfield  {author} {\bibinfo {author} {\bibfnamefont {M.~A.}\ \bibnamefont
			{Bandres}}, \bibinfo {author} {\bibfnamefont {S.}~\bibnamefont {Wittek}},
		\bibinfo {author} {\bibfnamefont {G.}~\bibnamefont {Harari}}, \bibinfo
		{author} {\bibfnamefont {M.}~\bibnamefont {Parto}}, \bibinfo {author}
		{\bibfnamefont {J.}~\bibnamefont {Ren}}, \bibinfo {author} {\bibfnamefont
			{M.}~\bibnamefont {Segev}}, \bibinfo {author} {\bibfnamefont {D.~N.}\
			\bibnamefont {Christodoulides}}, \ and\ \bibinfo {author} {\bibfnamefont
			{M.}~\bibnamefont {Khajavikhan}},\ }\href {\doibase 10.1126/science.aar4005}
	{\bibfield  {journal} {\bibinfo  {journal} {Science}\ } (\bibinfo {year}
		{2018}),\ 10.1126/science.aar4005}\BibitemShut {NoStop}%
	\bibitem [{\citenamefont {Harari}\ \emph {et~al.}(2018)\citenamefont {Harari},
		\citenamefont {Bandres}, \citenamefont {Lumer}, \citenamefont {Rechtsman},
		\citenamefont {Chong}, \citenamefont {Khajavikhan}, \citenamefont
		{Christodoulides},\ and\ \citenamefont {Segev}}]{Hararieaar4003}%
	\BibitemOpen
	\bibfield  {author} {\bibinfo {author} {\bibfnamefont {G.}~\bibnamefont
			{Harari}}, \bibinfo {author} {\bibfnamefont {M.~A.}\ \bibnamefont {Bandres}},
		\bibinfo {author} {\bibfnamefont {Y.}~\bibnamefont {Lumer}}, \bibinfo
		{author} {\bibfnamefont {M.~C.}\ \bibnamefont {Rechtsman}}, \bibinfo {author}
		{\bibfnamefont {Y.~D.}\ \bibnamefont {Chong}}, \bibinfo {author}
		{\bibfnamefont {M.}~\bibnamefont {Khajavikhan}}, \bibinfo {author}
		{\bibfnamefont {D.~N.}\ \bibnamefont {Christodoulides}}, \ and\ \bibinfo
		{author} {\bibfnamefont {M.}~\bibnamefont {Segev}},\ }\href {\doibase
		10.1126/science.aar4003} {\bibfield  {journal} {\bibinfo  {journal}
			{Science}\ } (\bibinfo {year} {2018}),\ 10.1126/science.aar4003}\BibitemShut
	{NoStop}%
	\bibitem [{\citenamefont {St-Jean}\ \emph {et~al.}(2017)\citenamefont
		{St-Jean}, \citenamefont {Goblot}, \citenamefont {Galopin}, \citenamefont
		{Lemaître}, \citenamefont {Ozawa}, \citenamefont {Le~Gratiet}, \citenamefont
		{Sagnes}, \citenamefont {Bloch},\ and\ \citenamefont {Amo}}]{amo2017}%
	\BibitemOpen
	\bibfield  {author} {\bibinfo {author} {\bibfnamefont {P.}~\bibnamefont
			{St-Jean}}, \bibinfo {author} {\bibfnamefont {V.}~\bibnamefont {Goblot}},
		\bibinfo {author} {\bibfnamefont {E.}~\bibnamefont {Galopin}}, \bibinfo
		{author} {\bibfnamefont {A.}~\bibnamefont {Lemaître}}, \bibinfo {author}
		{\bibfnamefont {T.}~\bibnamefont {Ozawa}}, \bibinfo {author} {\bibfnamefont
			{L.}~\bibnamefont {Le~Gratiet}}, \bibinfo {author} {\bibfnamefont
			{I.}~\bibnamefont {Sagnes}}, \bibinfo {author} {\bibfnamefont
			{J.}~\bibnamefont {Bloch}}, \ and\ \bibinfo {author} {\bibfnamefont
			{A.}~\bibnamefont {Amo}},\ }\href {https://doi.org/10.1038/s41566-017-0006-2}
	{\bibfield  {journal} {\bibinfo  {journal} {Nature Photonics}\ }\textbf
		{\bibinfo {volume} {11}},\ \bibinfo {pages} {651} (\bibinfo {year}
		{2017})}\BibitemShut {NoStop}%
	\bibitem [{\citenamefont {Rivolta}\ \emph {et~al.}(2017)\citenamefont
		{Rivolta}, \citenamefont {Benisty},\ and\ \citenamefont {Maes}}]{Rivolta}%
	\BibitemOpen
	\bibfield  {author} {\bibinfo {author} {\bibfnamefont {N.~X.~A.}\
			\bibnamefont {Rivolta}}, \bibinfo {author} {\bibfnamefont {H.}~\bibnamefont
			{Benisty}}, \ and\ \bibinfo {author} {\bibfnamefont {B.}~\bibnamefont
			{Maes}},\ }\href@noop {} {\bibfield  {journal} {\bibinfo  {journal} {Phys.
				Rev. A}\ }\textbf {\bibinfo {volume} {96}},\ \bibinfo {pages} {023864}
		(\bibinfo {year} {2017})}\BibitemShut {NoStop}%
	\bibitem [{\citenamefont {Pilozzi}\ and\ \citenamefont {Conti}(2016)}]{pilo}%
	\BibitemOpen
	\bibfield  {author} {\bibinfo {author} {\bibfnamefont {L.}~\bibnamefont
			{Pilozzi}}\ and\ \bibinfo {author} {\bibfnamefont {C.}~\bibnamefont
			{Conti}},\ }\href@noop {} {\bibfield  {journal} {\bibinfo  {journal} {Phys.
				Rev. B}\ }\textbf {\bibinfo {volume} {93}},\ \bibinfo {pages} {195317}
		(\bibinfo {year} {2016})}\BibitemShut {NoStop}%
	\bibitem [{\citenamefont {Zilberberg}\ \emph {et~al.}(2018)\citenamefont
		{Zilberberg}, \citenamefont {Huang}, \citenamefont {Guglielmon},
		\citenamefont {Wang}, \citenamefont {Chen}, \citenamefont {Kraus},\ and\
		\citenamefont {Rechtsman}}]{zilberberg_photonic_2018}%
	\BibitemOpen
	\bibfield  {author} {\bibinfo {author} {\bibfnamefont {O.}~\bibnamefont
			{Zilberberg}}, \bibinfo {author} {\bibfnamefont {S.}~\bibnamefont {Huang}},
		\bibinfo {author} {\bibfnamefont {J.}~\bibnamefont {Guglielmon}}, \bibinfo
		{author} {\bibfnamefont {M.}~\bibnamefont {Wang}}, \bibinfo {author}
		{\bibfnamefont {K.~P.}\ \bibnamefont {Chen}}, \bibinfo {author}
		{\bibfnamefont {Y.~E.}\ \bibnamefont {Kraus}}, \ and\ \bibinfo {author}
		{\bibfnamefont {M.~C.}\ \bibnamefont {Rechtsman}},\ }\href
	{http://dx.doi.org/10.1038/nature25011} {\bibfield  {journal} {\bibinfo
			{journal} {Nature}\ }\textbf {\bibinfo {volume} {553}},\ \bibinfo {pages}
		{59} (\bibinfo {year} {2018})}\BibitemShut {NoStop}%
	\bibitem [{\citenamefont {Lohse}\ \emph {et~al.}(2018)\citenamefont {Lohse},
		\citenamefont {Schweizer}, \citenamefont {Price}, \citenamefont
		{Zilberberg},\ and\ \citenamefont {Bloch}}]{lohse_exploring_2018}%
	\BibitemOpen
	\bibfield  {author} {\bibinfo {author} {\bibfnamefont {M.}~\bibnamefont
			{Lohse}}, \bibinfo {author} {\bibfnamefont {C.}~\bibnamefont {Schweizer}},
		\bibinfo {author} {\bibfnamefont {H.~M.}\ \bibnamefont {Price}}, \bibinfo
		{author} {\bibfnamefont {O.}~\bibnamefont {Zilberberg}}, \ and\ \bibinfo
		{author} {\bibfnamefont {I.}~\bibnamefont {Bloch}},\ }\href
	{http://dx.doi.org/10.1038/nature25000} {\bibfield  {journal} {\bibinfo
			{journal} {Nature}\ }\textbf {\bibinfo {volume} {553}},\ \bibinfo {pages}
		{55} (\bibinfo {year} {2018})}\BibitemShut {NoStop}%
	\bibitem [{\citenamefont {Zdeborova}(2017)}]{zde}%
	\BibitemOpen
	\bibfield  {author} {\bibinfo {author} {\bibfnamefont {L.}~\bibnamefont
			{Zdeborova}},\ }\href@noop {} {\bibfield  {journal} {\bibinfo  {journal}
			{Nat. Phys.}\ }\textbf {\bibinfo {volume} {13}},\ \bibinfo {pages} {420}
		(\bibinfo {year} {2017})}\BibitemShut {NoStop}%
	\bibitem [{\citenamefont {Carrasquilla}\ and\ \citenamefont
		{Melko}(2017)}]{carra}%
	\BibitemOpen
	\bibfield  {author} {\bibinfo {author} {\bibfnamefont {J.}~\bibnamefont
			{Carrasquilla}}\ and\ \bibinfo {author} {\bibfnamefont {R.~G.}\ \bibnamefont
			{Melko}},\ }\href@noop {} {\bibfield  {journal} {\bibinfo  {journal} {Nat.
				Phys.}\ }\textbf {\bibinfo {volume} {13}},\ \bibinfo {pages} {431} (\bibinfo
		{year} {2017})}\BibitemShut {NoStop}%
	\bibitem [{\citenamefont {Zhang}\ and\ \citenamefont {Kim}(2017)}]{zhang}%
	\BibitemOpen
	\bibfield  {author} {\bibinfo {author} {\bibfnamefont {Y.}~\bibnamefont
			{Zhang}}\ and\ \bibinfo {author} {\bibfnamefont {E.-A.}\ \bibnamefont
			{Kim}},\ }\href@noop {} {\bibfield  {journal} {\bibinfo  {journal} {Phys.
				Rev. Lett.}\ }\textbf {\bibinfo {volume} {118}},\ \bibinfo {pages} {216401}
		(\bibinfo {year} {2017})}\BibitemShut {NoStop}%
	\bibitem [{\citenamefont {Kraus}\ and\ \citenamefont
		{Zilberberg}(2012)}]{Kraus}%
	\BibitemOpen
	\bibfield  {author} {\bibinfo {author} {\bibfnamefont {Y.~E.}\ \bibnamefont
			{Kraus}}\ and\ \bibinfo {author} {\bibfnamefont {O.}~\bibnamefont
			{Zilberberg}},\ }\href@noop {} {\bibfield  {journal} {\bibinfo  {journal}
			{Phys. Rev. Lett.}\ }\textbf {\bibinfo {volume} {109}},\ \bibinfo {pages}
		{116404} (\bibinfo {year} {2012})}\BibitemShut {NoStop}%
	\bibitem [{\citenamefont {Aubry}\ and\ \citenamefont {Andr\'e}(1980)}]{Aubry}%
	\BibitemOpen
	\bibfield  {author} {\bibinfo {author} {\bibfnamefont {S.}~\bibnamefont
			{Aubry}}\ and\ \bibinfo {author} {\bibfnamefont {G.}~\bibnamefont
			{Andr\'e}},\ }\href@noop {} {\bibfield  {journal} {\bibinfo  {journal} {Ann.
				Israel. Phys. Soc.}\ }\textbf {\bibinfo {volume} {3}},\ \bibinfo {pages}
		{133} (\bibinfo {year} {1980})}\BibitemShut {NoStop}%
	\bibitem [{\citenamefont {Harper}(1955)}]{Harper}%
	\BibitemOpen
	\bibfield  {author} {\bibinfo {author} {\bibfnamefont {P.~G.}\ \bibnamefont
			{Harper}},\ }\href@noop {} {\bibfield  {journal} {\bibinfo  {journal} {Proc.
				Phys. Soc., London, Sect. A}\ }\textbf {\bibinfo {volume} {68}},\ \bibinfo
		{pages} {874} (\bibinfo {year} {1955})}\BibitemShut {NoStop}%
	\bibitem [{\citenamefont {Poshakinskiy}\ \emph {et~al.}(2014)\citenamefont
		{Poshakinskiy}, \citenamefont {Poddubny}, \citenamefont {Pilozzi},\ and\
		\citenamefont {Ivchenko}}]{Posha}%
	\BibitemOpen
	\bibfield  {author} {\bibinfo {author} {\bibfnamefont {A.~V.}\ \bibnamefont
			{Poshakinskiy}}, \bibinfo {author} {\bibfnamefont {A.~N.}\ \bibnamefont
			{Poddubny}}, \bibinfo {author} {\bibfnamefont {L.}~\bibnamefont {Pilozzi}}, \
		and\ \bibinfo {author} {\bibfnamefont {E.~L.}\ \bibnamefont {Ivchenko}},\
	}\href {\doibase 10.1103/PhysRevLett.112.107403} {\bibfield  {journal}
		{\bibinfo  {journal} {Phys. Rev. Lett.}\ }\textbf {\bibinfo {volume} {112}},\
		\bibinfo {pages} {107403} (\bibinfo {year} {2014})}\BibitemShut {NoStop}%
	\bibitem [{\citenamefont {Chew}(1999)}]{Chew}%
	\BibitemOpen
	\bibfield  {author} {\bibinfo {author} {\bibfnamefont {W.~C.}\ \bibnamefont
			{Chew}},\ }\href@noop {} {\emph {\bibinfo {title} {Waves and Fields in
				Inhomogeneous Media}}}\ (\bibinfo  {publisher} {Wiley-IEEE Press},\ \bibinfo
	{year} {1999})\BibitemShut {NoStop}%
	\bibitem [{\citenamefont {Hofstadter}(1976)}]{Hofs}%
	\BibitemOpen
	\bibfield  {author} {\bibinfo {author} {\bibfnamefont {D.~R.}\ \bibnamefont
			{Hofstadter}},\ }\href {\doibase 10.1103/PhysRevB.14.2239} {\bibfield
		{journal} {\bibinfo  {journal} {Phys. Rev. B}\ }\textbf {\bibinfo {volume}
			{14}},\ \bibinfo {pages} {2239} (\bibinfo {year} {1976})}\BibitemShut
	{NoStop}%
	\bibitem [{\citenamefont {Hatsugai}(1993)}]{Hatsugai}%
	\BibitemOpen
	\bibfield  {author} {\bibinfo {author} {\bibfnamefont {Y.}~\bibnamefont
			{Hatsugai}},\ }\href@noop {} {\bibfield  {journal} {\bibinfo  {journal}
			{Phys. Rev. B}\ }\textbf {\bibinfo {volume} {48}},\ \bibinfo {pages} {11851}
		(\bibinfo {year} {1993})}\BibitemShut {NoStop}%
	\bibitem [{\citenamefont {Tauber}\ and\ \citenamefont
		{Delplace}(2015)}]{Tauber}%
	\BibitemOpen
	\bibfield  {author} {\bibinfo {author} {\bibfnamefont {C.}~\bibnamefont
			{Tauber}}\ and\ \bibinfo {author} {\bibfnamefont {P.}~\bibnamefont
			{Delplace}},\ }\href@noop {} {\bibfield  {journal} {\bibinfo  {journal} {New
				J. Phys.}\ }\textbf {\bibinfo {volume} {17}},\ \bibinfo {pages} {115008}
		(\bibinfo {year} {2015})}\BibitemShut {NoStop}%
	\bibitem [{\citenamefont {Kabir}\ \emph {et~al.}(2008)\citenamefont {Kabir},
		\citenamefont {Wang}, \citenamefont {Yu},\ and\ \citenamefont
		{Zhang}}]{kabir}%
	\BibitemOpen
	\bibfield  {author} {\bibinfo {author} {\bibfnamefont {H.}~\bibnamefont
			{Kabir}}, \bibinfo {author} {\bibfnamefont {Y.}~\bibnamefont {Wang}},
		\bibinfo {author} {\bibfnamefont {M.}~\bibnamefont {Yu}}, \ and\ \bibinfo
		{author} {\bibfnamefont {Q.}~\bibnamefont {Zhang}},\ }\href@noop {}
	{\bibfield  {journal} {\bibinfo  {journal} {IEEE}\ }\textbf {\bibinfo
			{volume} {56}},\ \bibinfo {pages} {867} (\bibinfo {year} {2008})}\BibitemShut
	{NoStop}%
	\bibitem [{\citenamefont {Gosal}\ \emph {et~al.}(2016)\citenamefont {Gosal},
		\citenamefont {Almajali}, \citenamefont {McNamara},\ and\ \citenamefont
		{Yagoub}}]{Gosal}%
	\BibitemOpen
	\bibfield  {author} {\bibinfo {author} {\bibfnamefont {G.}~\bibnamefont
			{Gosal}}, \bibinfo {author} {\bibfnamefont {E.}~\bibnamefont {Almajali}},
		\bibinfo {author} {\bibfnamefont {D.}~\bibnamefont {McNamara}}, \ and\
		\bibinfo {author} {\bibfnamefont {M.}~\bibnamefont {Yagoub}},\ }\href@noop {}
	{\bibfield  {journal} {\bibinfo  {journal} {IEEE, Antennas and wireless
				propagation letters}\ }\textbf {\bibinfo {volume} {15}},\ \bibinfo {pages}
		{1483} (\bibinfo {year} {2016})}\BibitemShut {NoStop}%
	\bibitem [{\citenamefont {Aoad}\ \emph {et~al.}(2017)\citenamefont {Aoad},
		\citenamefont {Simsek},\ and\ \citenamefont {Aydin}}]{Aoad}%
	\BibitemOpen
	\bibfield  {author} {\bibinfo {author} {\bibfnamefont {A.}~\bibnamefont
			{Aoad}}, \bibinfo {author} {\bibfnamefont {M.}~\bibnamefont {Simsek}}, \ and\
		\bibinfo {author} {\bibfnamefont {Z.}~\bibnamefont {Aydin}},\ }\href
	{\doibase 10.1002/jnm.2129} {\bibfield  {journal} {\bibinfo  {journal}
			{International Journal of Numerical Modelling: Electronic Networks, Devices
				and Fields}\ }\textbf {\bibinfo {volume} {30}},\ \bibinfo {pages} {e2129}
		(\bibinfo {year} {2017})},\ \bibinfo {note} {e2129
		JNM-15-0044.R1}\BibitemShut {NoStop}%
	\bibitem [{\citenamefont {Kartashov}\ and\ \citenamefont
		{Skryabin}(2017)}]{Kartashov}%
	\BibitemOpen
	\bibfield  {author} {\bibinfo {author} {\bibfnamefont {Y.~V.}\ \bibnamefont
			{Kartashov}}\ and\ \bibinfo {author} {\bibfnamefont {D.~V.}\ \bibnamefont
			{Skryabin}},\ }\href {\doibase 10.1103/PhysRevLett.119.253904} {\bibfield
		{journal} {\bibinfo  {journal} {Phys. Rev. Lett.}\ }\textbf {\bibinfo
			{volume} {119}},\ \bibinfo {pages} {253904} (\bibinfo {year}
		{2017})}\BibitemShut {NoStop}%
	\bibitem [{\citenamefont {Mihalache}\ \emph {et~al.}(2010)\citenamefont
		{Mihalache}, \citenamefont {Mazilu}, \citenamefont {Skarka}, \citenamefont
		{Malomed}, \citenamefont {Leblond}, \citenamefont {Aleksi\'c},\ and\
		\citenamefont {Lederer}}]{Mihalache}%
	\BibitemOpen
	\bibfield  {author} {\bibinfo {author} {\bibfnamefont {D.}~\bibnamefont
			{Mihalache}}, \bibinfo {author} {\bibfnamefont {D.}~\bibnamefont {Mazilu}},
		\bibinfo {author} {\bibfnamefont {V.}~\bibnamefont {Skarka}}, \bibinfo
		{author} {\bibfnamefont {B.~A.}\ \bibnamefont {Malomed}}, \bibinfo {author}
		{\bibfnamefont {H.}~\bibnamefont {Leblond}}, \bibinfo {author} {\bibfnamefont
			{N.~B.}\ \bibnamefont {Aleksi\'c}}, \ and\ \bibinfo {author} {\bibfnamefont
			{F.}~\bibnamefont {Lederer}},\ }\href@noop {} {\bibfield  {journal} {\bibinfo
			{journal} {Phys. Rev. A}\ }\textbf {\bibinfo {volume} {82}},\ \bibinfo
		{pages} {023813} (\bibinfo {year} {2010})}\BibitemShut {NoStop}%
	\bibitem [{\citenamefont {J\"unemann}\ \emph {et~al.}(2017)\citenamefont
		{J\"unemann}, \citenamefont {Piga}, \citenamefont {Ran}, \citenamefont
		{Lewenstein}, \citenamefont {Rizzi},\ and\ \citenamefont
		{Bermudez}}]{PhysRevX.7.031057}%
	\BibitemOpen
	\bibfield  {author} {\bibinfo {author} {\bibfnamefont {J.}~\bibnamefont
			{J\"unemann}}, \bibinfo {author} {\bibfnamefont {A.}~\bibnamefont {Piga}},
		\bibinfo {author} {\bibfnamefont {S.-J.}\ \bibnamefont {Ran}}, \bibinfo
		{author} {\bibfnamefont {M.}~\bibnamefont {Lewenstein}}, \bibinfo {author}
		{\bibfnamefont {M.}~\bibnamefont {Rizzi}}, \ and\ \bibinfo {author}
		{\bibfnamefont {A.}~\bibnamefont {Bermudez}},\ }\href {\doibase
		10.1103/PhysRevX.7.031057} {\bibfield  {journal} {\bibinfo  {journal} {Phys.
				Rev. X}\ }\textbf {\bibinfo {volume} {7}},\ \bibinfo {pages} {031057}
		(\bibinfo {year} {2017})}\BibitemShut {NoStop}%
	\bibitem [{\citenamefont {Mancini}\ \emph {et~al.}(2015)\citenamefont
		{Mancini}, \citenamefont {Pagano}, \citenamefont {Cappellini}, \citenamefont
		{Livi}, \citenamefont {Rider}, \citenamefont {Catani}, \citenamefont {Sias},
		\citenamefont {Zoller}, \citenamefont {Inguscio}, \citenamefont {Dalmonte},\
		and\ \citenamefont {Fallani}}]{Mancini1510}%
	\BibitemOpen
	\bibfield  {author} {\bibinfo {author} {\bibfnamefont {M.}~\bibnamefont
			{Mancini}}, \bibinfo {author} {\bibfnamefont {G.}~\bibnamefont {Pagano}},
		\bibinfo {author} {\bibfnamefont {G.}~\bibnamefont {Cappellini}}, \bibinfo
		{author} {\bibfnamefont {L.}~\bibnamefont {Livi}}, \bibinfo {author}
		{\bibfnamefont {M.}~\bibnamefont {Rider}}, \bibinfo {author} {\bibfnamefont
			{J.}~\bibnamefont {Catani}}, \bibinfo {author} {\bibfnamefont
			{C.}~\bibnamefont {Sias}}, \bibinfo {author} {\bibfnamefont {P.}~\bibnamefont
			{Zoller}}, \bibinfo {author} {\bibfnamefont {M.}~\bibnamefont {Inguscio}},
		\bibinfo {author} {\bibfnamefont {M.}~\bibnamefont {Dalmonte}}, \ and\
		\bibinfo {author} {\bibfnamefont {L.}~\bibnamefont {Fallani}},\ }\href
	{\doibase 10.1126/science.aaa8736} {\bibfield  {journal} {\bibinfo  {journal}
			{Science}\ }\textbf {\bibinfo {volume} {349}},\ \bibinfo {pages} {1510}
		(\bibinfo {year} {2015})}\BibitemShut {NoStop}%
	\bibitem [{\citenamefont {Parto}\ \emph {et~al.}()\citenamefont {Parto},
		\citenamefont {Wittek}, \citenamefont {Hodaei}, \citenamefont {Harari},
		\citenamefont {Bandres}, \citenamefont {Ren}, \citenamefont {Rechtsman},
		\citenamefont {Segev}, \citenamefont {Christodoulides},\ and\ \citenamefont
		{Khajavikhan}}]{PartoSegev}%
	\BibitemOpen
	\bibfield  {author} {\bibinfo {author} {\bibfnamefont {M.}~\bibnamefont
			{Parto}}, \bibinfo {author} {\bibfnamefont {S.}~\bibnamefont {Wittek}},
		\bibinfo {author} {\bibfnamefont {H.}~\bibnamefont {Hodaei}}, \bibinfo
		{author} {\bibfnamefont {G.}~\bibnamefont {Harari}}, \bibinfo {author}
		{\bibfnamefont {M.}~\bibnamefont {Bandres}}, \bibinfo {author} {\bibfnamefont
			{J.}~\bibnamefont {Ren}}, \bibinfo {author} {\bibfnamefont {M.}~\bibnamefont
			{Rechtsman}}, \bibinfo {author} {\bibfnamefont {M.}~\bibnamefont {Segev}},
		\bibinfo {author} {\bibfnamefont {D.}~\bibnamefont {Christodoulides}}, \ and\
		\bibinfo {author} {\bibfnamefont {M.}~\bibnamefont {Khajavikhan}},\
	}\href@noop {} {\bibinfo  {journal} {arXiv:1709.00523}\ }\BibitemShut
	{NoStop}%
	\bibitem [{\citenamefont {Pilozzi}\ and\ \citenamefont
		{C.Conti}(2017)}]{Laura}%
	\BibitemOpen
	\bibfield  {journal} {  }\bibfield  {author} {\bibinfo {author} {\bibfnamefont
			{L.}~\bibnamefont {Pilozzi}}\ and\ \bibinfo {author} {\bibnamefont
			{C.Conti}},\ }\href@noop {} {\bibfield  {journal} {\bibinfo  {journal} {Opt.
				Lett.}\ }\textbf {\bibinfo {volume} {42}},\ \bibinfo {pages} {5174} (\bibinfo
		{year} {2017})}\BibitemShut {NoStop}%
	\bibitem [{\citenamefont {{Longhi}}(2018)}]{Longhi2018}%
	\BibitemOpen
	\bibfield  {author} {\bibinfo {author} {\bibfnamefont {S.}~\bibnamefont
			{{Longhi}}},\ }\href@noop {} {\bibfield  {journal} {\bibinfo  {journal}
			{ArXiv e-prints}\ } (\bibinfo {year} {2018})},\ \Eprint
	{http://arxiv.org/abs/1802.05025} {arXiv:1802.05025 [physics.optics]}
	\BibitemShut {NoStop}%
	\bibitem [{\citenamefont {Zeuner}\ \emph {et~al.}(2015)\citenamefont {Zeuner},
		\citenamefont {Rechtsman}, \citenamefont {Plotnik}, \citenamefont {Lumer},
		\citenamefont {Nolte}, \citenamefont {Rudner}, \citenamefont {Segev},\ and\
		\citenamefont {Szameit}}]{Zeuner}%
	\BibitemOpen
	\bibfield  {author} {\bibinfo {author} {\bibfnamefont {J.~M.}\ \bibnamefont
			{Zeuner}}, \bibinfo {author} {\bibfnamefont {M.~C.}\ \bibnamefont
			{Rechtsman}}, \bibinfo {author} {\bibfnamefont {Y.}~\bibnamefont {Plotnik}},
		\bibinfo {author} {\bibfnamefont {Y.}~\bibnamefont {Lumer}}, \bibinfo
		{author} {\bibfnamefont {S.}~\bibnamefont {Nolte}}, \bibinfo {author}
		{\bibfnamefont {M.~S.}\ \bibnamefont {Rudner}}, \bibinfo {author}
		{\bibfnamefont {M.}~\bibnamefont {Segev}}, \ and\ \bibinfo {author}
		{\bibfnamefont {A.}~\bibnamefont {Szameit}},\ }\href {\doibase
		10.1103/PhysRevLett.115.040402} {\bibfield  {journal} {\bibinfo  {journal}
			{Phys. Rev. Lett.}\ }\textbf {\bibinfo {volume} {115}},\ \bibinfo {pages}
		{040402} (\bibinfo {year} {2015})}\BibitemShut {NoStop}%
	\bibitem [{\citenamefont {Graf}\ and\ \citenamefont {Porta}(2013)}]{Graf}%
	\BibitemOpen
	\bibfield  {author} {\bibinfo {author} {\bibfnamefont {G.}~\bibnamefont
			{Graf}}\ and\ \bibinfo {author} {\bibfnamefont {M.}~\bibnamefont {Porta}},\
	}\href@noop {} {\bibfield  {journal} {\bibinfo  {journal} {Commun. Math.
				Phys.}\ }\textbf {\bibinfo {volume} {324}},\ \bibinfo {pages} {851} (\bibinfo
		{year} {2013})}\BibitemShut {NoStop}%
	\bibitem [{\citenamefont {Poshakinskiy}\ \emph {et~al.}(2015)\citenamefont
		{Poshakinskiy}, \citenamefont {Poddubny},\ and\ \citenamefont
		{Hafezi}}]{PoshaAr}%
	\BibitemOpen
	\bibfield  {author} {\bibinfo {author} {\bibfnamefont {A.~V.}\ \bibnamefont
			{Poshakinskiy}}, \bibinfo {author} {\bibfnamefont {A.~N.}\ \bibnamefont
			{Poddubny}}, \ and\ \bibinfo {author} {\bibfnamefont {M.}~\bibnamefont
			{Hafezi}},\ }\href {\doibase 10.1103/PhysRevA.91.043830} {\bibfield
		{journal} {\bibinfo  {journal} {Phys. Rev. A}\ }\textbf {\bibinfo {volume}
			{91}},\ \bibinfo {pages} {043830} (\bibinfo {year} {2015})}\BibitemShut
	{NoStop}%
\end{thebibliography}
%merlin.mbs apsrev4-1.bst 2010-07-25 4.21a (PWD, AO, DPC) hacked
%Control: key (0)
%Control: author (8) initials jnrlst
%Control: editor formatted (1) identically to author
%Control: production of article title (-1) disabled
%Control: page (0) single
%Control: year (1) truncated
%Control: production of eprint (0) enabled
%

\end{document}